\documentclass[aps, twocolumn, superscriptaddress, showpacs, nofootinbib, longbibliography]{revtex4-1}

\usepackage[utf8]{inputenc}
\usepackage[T1]{fontenc}
\usepackage{ae,aecompl} 
\usepackage{graphicx}
\usepackage{amsmath}
\usepackage{xcolor}
\usepackage{amssymb}
\usepackage{latexsym}
\usepackage{wasysym}
\usepackage{psfrag}
\usepackage{ifthen}
\usepackage[citecolor=blue,colorlinks=true]{hyperref}
\usepackage{float}
\usepackage[utf8]{inputenc}
\usepackage{lineno}
\usepackage{multirow}
\usepackage{subcaption}
\usepackage{orcidlink}

\definecolor{blue-violet}{rgb}{0.54, 0.17, 0.89}

\begin{document}

\title{Enhancing search pipelines for short gravitational-wave transients with Gaussian mixture modeling}

\author{Leigh Smith \orcidlink{0000-0002-3035-0947}}
\affiliation{SUPA, School of Physics and Astronomy, University of Glasgow, Glasgow G12 8QQ, United Kingdom}
\affiliation{Dipartimento di Fisica, Università di Trieste, I-34127 Trieste, Italy}
\affiliation{INFN, Sezione di Trieste, I-34127 Trieste, Italy}
\author{Sayantan Ghosh \orcidlink{0009-0002-2884-6836}}
\affiliation{Department of Physics, Indian Institute of Technology Bombay, Mumbai, Maharashtra 400076, India}
\author{Jiyoon Sun \orcidlink{0009-0008-8278-0077}}
\affiliation{Department of Physics, Chung-Ang University, Seoul 06974, Korea}
\author{V. Gayathri \orcidlink{0000-0002-7167-9888}}
\affiliation{Leonard E. Parker Center for Gravitation, Cosmology, and Astrophysics, University of Wisconsin–Milwaukee, Milwaukee, WI 53201, USA}
\author{Ik Siong Heng \orcidlink{0000-0002-1977-0019}}
\affiliation{SUPA, School of Physics and Astronomy, University of Glasgow, Glasgow G12 8QQ, United Kingdom}
\author{Archana Pai \orcidlink{0000-0003-3476-4589}}
\affiliation{Department of Physics, Indian Institute of Technology Bombay, Mumbai, Maharashtra 400076, India}

\begin{abstract}

We present an enhanced method for the application of Gaussian Mixture Modelling (GMM) to the coherent WaveBurst (cWB) algorithm in the search for short-duration gravitational wave (GW) transients. The supervised Machine Learning method of GMM allows for the multi-dimensional distributions of noise and signal to be modelled over a set of representative attributes, which aids in the classification of GW signals against noise transients (glitches) in the data. We demonstrate that updating the approach to model construction eliminates bias previously seen in the GMM analysis, increasing the robustness and sensitivity of the analysis over a wider range of burst source populations. The enhanced methodology is applied to the generic burst all-sky short search in the LIGO-Virgo full third observing run (O3), marking the first application of GMM to the 3 detector Livingston-Hanford-Virgo network. For both 2- and 3- detector networks, we observe comparable sensitivities to an array of generic signal morphologies, with significant sensitivity improvements to waveforms in the low Quality factor parameter space at false alarm rates of 1 per 100 years. This proves that GMM can effectively mitigate blip glitches, which are one of the most problematic sources of noise for un-modelled GW searches. The cWB-GMM search recovers similar numbers of compact binary coalescence (CBC) events as other cWB post-production methods, and concludes on no new gravitational wave detection after known CBC events are removed.

\end{abstract}

\maketitle

\section{Introduction}
\label{Sec:intro}

Since the first detection of gravitational waves (GWs) in 2015 \cite{LIGOScientific:GW150914}, the LIGO-Virgo-KAGRA (LVK) collaboration have detected approximately 90 gravitational wave events over three observing runs \cite{GWTC-1,GWTC-2,GWTC-2.1,GWTC-3}. These signals have all occurred due to compact binary coalescences (CBC) involving black holes and neutron stars, with detections including binary black hole mergers, binary neutron star systems \cite{gw170817-discovery,gw170817-properties,gw190425-discovery}, neutron star - black hole systems \cite{2nsbh-discoveries}, and even the possibility of an intermediate mass black hole system \cite{gw190521-discovery,gw190521-properties,o3-imbh}. There are other potential sources of gravitational waves that have yet to be discovered. One such source are generic GW transients ({\it Bursts}). The search for burst signals from short duration GW transients, known as the all-sky short search, aims to detect GW events with duration upto 10 seconds in the advanced ground based detectors. Some of the predicted sources in this search are core-collapse supernovae (CCSN) \cite{Powell:2018-s18,Powell:2020-m39,Radice:2018-s9,OConnor:2018-m20,Abdikamalov:2020jzn,CCSN-o3-Szczepanczyk}, cosmic strings \cite{CosmicString-o1-LVK,CosmicString-o3-LVK}, hyperbolic black hole encounters \cite{Cho:2018hyp,Morras:2021hypgw,Dandapat:2023hyp,Bini:2023hyp}, radiation driven parabolic capture \cite{Bae:2017capture,Ebersold:2022cap}, non-linear memory effects \cite{NL-memory-ebersold20,Hubner:2019NL-mem}, and neutron star glitches \cite{NSgkitch-2006-LVK,Lopez:2022NS-glitch,Yim:2020NS-glitch}. Many of the astrophysical sources mentioned do not have well-known waveform structure or have waveforms which are too computationally expensive to use in typical matched filtering searches \cite{LIGOScientific:matchedfilter}. This, along with the requirement of being sensitive to a wide variety of sources, requires burst searches to employ un-modelled techniques with little assumptions made on source waveforms. Such techniques are crucial for ensuring that any source of GW signals are not overlooked, however also present challenges in differentiating such events from detector noise transients.

One particular burst analysis algorithm is coherent WaveBurst (cWB) \cite{Klimenko_2004wavelet,Klimenko_2008cwb,Klimenko:2015cwb,Drago:2020cwb}, which is one of the main pipelines used to search for short duration transients in previous LVK observing runs \cite{allsky-o1,allsky-o2,allskyo3}, and has also contributed to the detection of CBC events in previous gravitational wave transient catalogs \cite{GWTC-1,GWTC-2,GWTC-3}. cWB bases the detection of a GW signal on excess coherent power within a network of GW detectors, assuming very little on signal morphology. Because of this, the algorithm is susceptible to non-stationary noise transients in the detector data, commonly referred to as `glitches'. Of particular concern are blip glitches \cite{Cabero:2019blip} which mimic the morphology of short-duration astrophysical signals, thereby hindering the optimisation of search sensitivity. Previously it was attempted to overcome this manually through visual inspection of the algorithm's response to simulated signals, however it has recently been thought that Machine Learning (ML) approaches can be utilized to distinguish GW signals from noisy glitches in a more efficient way. In this paper we will focus on the supervised ML method of Gaussian Mixture Modelling (GMM) as a post-production to the cWB algorithm. 

The GMM methodology models cWB attributes in multi-dimensional space as a superposition of Gaussians, allowing for the signal and noise populations to be modelled separately and thus aiding in the classification of GW signals while preserving the un-modelled nature of the search algorithm. The benefit of this has been outlined by previous studies in \cite{Gayathri_2020,Lopez_2022}, however in this work we present new alterations to the GMM post-production methodology, including new training data, updated trigger attributes and approach for obtaining optimal number of Gaussians in the models. Through this, we remove a bias that was previously present within the models due to the choice of signal model training data, thus improving the robustness of our analysis to the wide GW burst parameter space. We focus particularly on the sensitivities for the all-sky short search in the third observing run (O3), comparing sensitivities from the first half of the third observing run (O3a) to previous GMM studies in \cite{Lopez_2022}, and presenting results for the full O3 search with GMM post-production for the first time with both 2- and 3-detector networks.

Recent work has seen the application of a different ML-based approach as post-production to the cWB algorithm \cite{xgboost_2023}, in which the authors used the decision tree method, XGBoost, to improve the classification of signal and noise transients while maintaining un-modelled requirements to the analysis. XGBoost learns how to discriminate between typical noise and signal population features through chosen cWB summary statistics, outputting a number between 0 (noise) and 1 (signal) which weights the SNR of given events. As an alternative method to GMM post-production, comparisons of performance are made to the XGBoost approach throughout the paper.

%% STRUCTURE OF PAPER
The paper is structured as follows. Section \ref{Sec:Methodology} details the enhanced methodology of Gaussian Mixture Modelling and the application as post-production to the cWB algorithm. Section \ref{Sec:robustness_o3a} explores the robustness of the updated methodology through comparisons to results with the previous method in O3a data. Section \ref{Sec:results_o3b} gives an overview of the results on the LVK full O3 observing run using GMM post-production with cWB for 2- and 3-detector networks, including sensitivity to generic signal morphologies, core-collapse supernova waveforms, cosmic strings and detected GW events. Finally, in Section \ref{Sec:conclusions}, we summarise the key works from the GMM method and future plans.

\section{Methodology}
\label{Sec:Methodology}

\subsection{Coherent WaveBurst}

Coherent WaveBurst (cWB) \cite{Klimenko_2008cwb,Klimenko:2015cwb,Drago:2020cwb} is an un-modelled search algorithm which holds no assumption on a potential signal's morphology, sky direction or polarisation. Instead, a coherent analysis is used across multiple detectors, transforming time-domain strain data into the time-frequency domain via the Wilson-Daubechiers-Meyer wavelet transformation \cite{Necula_2012}. Pixels with excess coherent energy are selected above a given noise threshold from the network of detectors, while attributes are calculated based on statistics from signal and noise properties. These pixels are clustered based on time-frequency information with the help of the nearest neighbouring algorithm. From here clusters are reconstructed and are labelled as possible GW events ({\it triggers}) if they surpass thresholds on coherent energy ($E_{c}$) and network correlation coefficient ($c_{c}$).

Following the production of triggers, search sensitivities are optimised during a post-production stage, in which trigger attributes are manipulated in an effort to distinguish potential signals from noise transients. The standard cWB post-production method which was previously utilized in the LVK all-sky short search \cite{allskyo3} does this by placing threshold cuts on triggers' coherent statistics in an attempt to remove noisy glitches. Triggers are then split into 3 search classes based upon energy distribution and Quality factor, Q, so that problematic blip glitches are refined to only one class, and the significance of events in the other classes remains unaffected. While this methodology is effective, it does not distinguish triggers with low Q and single-energy oscillation well, such as Gaussian Pulses, since the bulk of problematic glitches lie within this class. Defining the attribute thresholds in which to define these search classes can also be a challenging and lengthy process. 
%\gayathri{details about 3 search classes and give reference}

\subsection{Gaussian Mixture Modelling}
\label{overview-GMM}
Gaussian Mixture Modelling (GMM) is a supervised Machine Learning method which allows for multi-model data to be modelled as superpositions of Gaussians. By constructing two distinct GMM models based on signal and noise distributions in the multi-dimensional attribute space, it is possible to distinguish astrophysical GW signals from noisy glitches by calculating likelihoods with respect to the models. The full methodology of using GMM to aid the search of gravitational waves was first proposed in \cite{Gayathri_2020}, while the application of GMM as a post-production to cWB was detailed for the all-sky short burst search during O3a in \cite{Lopez_2022}. The approach to constructing and optimising models has since been altered in order to enhance the sensitivity and robustness of the analysis to a wide range of GW burst waveform types. The updated approach is outlined in section \ref{gmm-cwb}. 

The GMM analysis is applied to triggers which have multi-dimensional statistical attributes which represent the distributions of noise and signal well. The triggers representing GW signals and detector noise can be considered as two distinct populations, allowing for two completely separate GMM models to be constructed as superpositions of Gaussians. Once these models are constructed, the log-likelihood of a given trigger being in either model can be determined through $W = \ln(\hat{\mathcal{L}} )|_{\hat{K}}$, where $\hat{\mathcal{L}}$ is the value of the likelihood function given an optimal number of $\hat{K}$ Gaussians \cite{Gayathri_2020}. The optimal number of Gaussians $\hat{K}$ are found by optimising the detection efficiency on validation data, explained in detail in section \ref{gmm-pp-cwb}. This leads to the construction of the GMM detection statistic, $T$: 
\begin{equation} \label{eq:T-defn}
    T = W_{s}- W_{n}
\end{equation}
where the subscripts ${s}$ and ${n}$ stand for signal and noise. $T$ is a log-likelihood ratio measure of a given trigger belonging to the distribution of GW signals, with positive $T$ values favouring signal and negative $T$ values favouring noise. The GMM methodology described above can be applied to any search in which the signal and noise parameter space are distinguishable over selected attributes, however we outline below the details specific to cWB application.

\subsection{Application to the all-sky short search with coherent WaveBurst}\label{gmm-cwb}

The following sections detail how GMM is applied as post-production to the cWB algorithm, replacing the standard methodology based upon binning in an attempt to mitigate the affect of blip glitches and better improve the classification of signals against noise from the detectors.

\subsubsection{Data}
The triggers and calculated attributes, derived from the cWB production, are used as inputs for the GMM analysis. A portion of these triggers is reserved for training and validating the models.
%The triggers and calculated attributes, the output from cWB production, are taken as inputs to the GMM analysis, with a portion of triggers reserved for training and validating the models. 
The background triggers represent a population of detector glitches, produced by cWB through simulated time-shifts so that it is un-physical for a GW signal to exist and only detector noise is considered. These background triggers are split into three subsets: validation, training and testing data. The training data is used to model the background GMM, while the validation data is used to tune the optimal number of Gaussian components per model, and test data is reserved for False Alarm Rate (FAR) calculation with the final GMM models.

To construct the signal model, generic band-limited white noise burst (WNB) injections are simulated to represent the wide range of signal attribute space, as in the XGBoost post-production in \cite{xgboost_2023}. The WNBs span the low-frequency range of the all-sky short search, and are designed to cover the signal parameter space over selected attributes. Specifically, it consists of two distributions: firstly WNBs are uniformly distributed between central frequency range $[24,996]$Hz and bandwidth [10,300]Hz, with duration logarithmically distributed between $[0.1,500]$ms; the second has WNBs with bandwidth of 10Hz, duration randomly distributed over $[0.1,10]$ms, and frequency over $[10,100]$Hz\footnote{for the secondary WNB distribution, frequency is dependent on duration$^{-0.5}$}. Unlike the background data, the WNB triggers are split into only two subsets of validation and training, with training data used to fit the GMM signal model and validation again used for the tuning of number of Gaussian components. The choice of generic simulated WNBs as training implies that the signal model is more representative of the entire burst sensitive parameter space, rather than training on a distribution that represents fewer samples in frequency, as was done in \cite{Lopez_2022}. It also builds a model that is relatively less biased towards any specific population of GW sources, since WNBs have random waveform morphologies over the frequency, bandwidth and duration parameters. Further, this choice of training data reserves all ad-hoc simulation injections for sensitivity estimates. 

For the multi-dimensional attribute space, we select a subset of cWB attributes in which properties of signals and noise are well represented, and re-parameterise to achieve desirable Gaussian behaviour. The attributes considered in the GMM analysis are as follows: effective network coherent SNR ($\eta_{c}$), network correlation coefficients ($c_{c0}$, $c_{c2}$), the network coherent energy ($E_{c}$), the network energy dis-balance ($N_{ED}$), the ratio between the reconstructed energy and the total energy ($N_{\mathrm{norm}}$), the penalty factor (penalty), and attributes measuring likeness to known glitches ($Q_{veto}$, $L_{veto}$). The re-parameterisation of these attributes are seen in Table \ref{tab:attributes}, with more details of their distributions in Appendix \ref{app:attributes}.
In the previous methodology \cite{Lopez_2022}, the $L_{veto2}$ attribute had also been considered, which aids in identifying narrow-band glitches observed at power line frequencies. However with the new training data outlined above, it was found that the consideration of this attribute was causing confusion to the models and that this source of glitches were not found to be problematic in O3 data. Hence, this attribute is no longer considered.

\begin{table}
    \centering
    \begin{tabular}{c c c}
         \hline
         \hline

          Original attribute & \multicolumn{2}{c}{Re-parameterised attribute}  \\ 
             &  LH &  LHV \\[1ex] 
          \hline
 
          $E_{c}$ &  \multicolumn{2}{c}{$\log_{10}(E_{c})$} \\
          $\eta_{c}$ &  \multicolumn{2}{c}{$\log_{10}(\eta_{c})$} \\ 
          $c_{c0}$ & \multicolumn{2}{c}{$\mathrm{logit}(c_{c0})$} \\
          $c_{c2}$ & \multicolumn{2}{c}{$\mathrm{logit}(c_{c2})$} \\
          $N_{ED}$ & $\log_{10}(N_{ED}+1000)$, & $\log_{10}(N_{ED}+2000)$ \\
          $N_{\mathrm{norm}}$ & \multicolumn{2}{c}{$N_{\mathrm{norm}}$} \\
          $\mathrm{penalty}$ & \multicolumn{2}{c}{$\log_{10}(\mathrm{penalty})$} \\ 
          $Q_{\mathrm{veto0}}$ & \multicolumn{2}{c}{$\log_{10}(Q_{\mathrm{veto0}}+1)$} \\
          $Q_{\mathrm{veto1}}$ & $\log_{10}(Q_{\mathrm{veto1}})$, & $\log_{10}(Q_{\mathrm{veto1}}+1)$ \\
          $L_{\mathrm{veto0}}$ & \multicolumn{2}{c}{\multirow{2}{*}{$\mathrm{logit}(L_{\mathrm{ratio}}) = \mathrm{logit}(\frac{L_{\mathrm{veto1}}}{L_{\mathrm{veto0}}})$}} \\
          $L_{\mathrm{veto1}}$ & \\[1ex] 
    \hline
    \hline
         
    \end{tabular}
    \caption{Table of cWB attributes selected for GMM analysis and their re-parameterisation. Re-parameterisation is similar for LH and LHV networks, with differences only in $N_{ED}$ and $Q_{\mathrm{veto1}}$. }
    \label{tab:attributes}
\end{table}

\subsubsection{Model optimisation}\label{gmm-pp-cwb}

\begin{figure}
    \centering
    \includegraphics[width=0.49\textwidth]{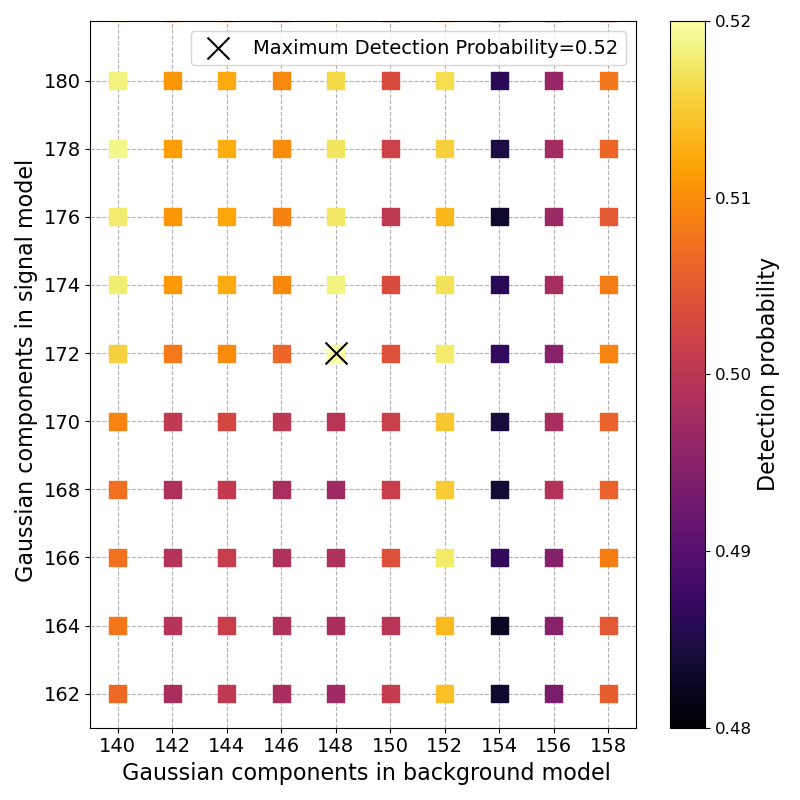}
    \caption{Variation of the detection efficiency with the number of Gaussian components. $N_{BKG}$ and $N_{SIM}$ are the numbers of Gaussian components for the background and the WNB distributions respectively. The colorbar shows the detection efficiency, which is the fraction of the validation WNB triggers detected at an FAR of 1 in 50 years or lower.}
    \label{fig:dp_valid_colorbar}
\end{figure}

As mentioned above, a subset of background triggers and simulated WNB triggers are used to construct the background and signal GMM models respectively. While fitting mixture models to each of the two training sets, Gaussian parameters (mean, covariance and weight of each Gaussian) are optimised by the Expectation Maximisation (EM) algorithm \cite{EM_1977}, while the hyper-parameter of number of Gaussian components is specified by optimising detection efficiency with the validation data. During this validation step, models are produced over a range of number of Gaussian components. The log-likelihood values of all triggers (WNB and background) in validation data are calculated for each background and signal model over the range of the number of Gaussian components, allowing for the construction of the $T$ statistic as in eqn. \ref{eq:T-defn}. We define detection efficiency as the fraction of WNB validation triggers with $T$ value greater than the detection threshold at a given FAR. In general, the fraction of detected signals will depend on the waveform type and the injected amplitude. But for the detection efficiency calculation here, we use only the WNB triggers and consider all triggers above the detection threshold regardless of the amplitude. For model optimisation, we choose to calculate the detection efficiency at a FAR of 1 in 50 years for each combination of GMM models. The combination of models which give the optimal detection efficiency at the selected FAR threshold are labelled as the GMM models with the optimal number of Gaussian components. An example of the distribution of efficiency at FAR 1 in 50 years over the number of Gaussian components is seen for the O3a analysis in fig \ref{fig:dp_valid_colorbar}. 
GMMs were trained for number of Gaussian components in the range $[140,168]$ for background triggers and in the range $[120,188]$ for the WNB simulations, both varying in steps of 2. Only the region surrounding the optimal number of  Gaussian components is shown in the figure. The colorbar represents the detection efficiency on validation data for each combination of models. We observe that the detection efficiency varies from $0.48$ to $0.52$ in this range of numbers of Gaussian components, with the maximum occurring at the combination of models with $148$ components for background and $172$ components for WNB simulations.

Previously in \cite{Lopez_2022}, the optimal number of Gaussian components had been selected using the Bayesian Information Criterion (BIC). However, the BIC provided a measure of how well the Gaussians fit the training data, which did not directly relate to our models' ability to distinguish between the background and signal populations. The new approach attempts to combine both measures through the optimisation of correctly classified data, and was found to reliably obtain better results.

Once the optimal models have been chosen, the $T$ detection statistic is calculated for each trigger in the background and signal test data sets, as detailed in section \ref{overview-GMM}, to assign triggers significance estimates.

\section{Robustness checks with O3a}
\label{Sec:robustness_o3a}

The GMM methodology detailed in section \ref{Sec:Methodology} ensures the analysis is sensitive to a variety of signals that may arise from the all-sky short search. Here, we investigate the robustness of this updated GMM methodology by comparing sensitivities to the previous GMM post-production in \cite{Lopez_2022} for O3a, from 1st April 2019 to 1st October 2019. We consider the 2-detector LIGO-Livingston-LIGO-Hanford (LH) network for the frequency range 16-1024 Hz, giving a total coincidence time of 104.9 days, while a total of 980.7 years of background data is available through the application of un-physical time-shifts between detector data.

The robustness of the analysis is investigated by making population injections into O3a data with cWB using the generic ad-hoc set of waveforms commonly used to benchmark 
%(\gayathri{do we want to use benchmarked or mock data challenge}) 
all-sky short pipeline sensitivities, and astrophysically motivated core-collapse supernovae (CCSN) waveforms. The set of generic ad-hoc simulations used in the burst all-sky short search \cite{allskyo3} consists of Gaussian Pulse (GA), Sine-Gaussian (SG) and White Noise Burst (WNB) simulations, injected over a variety of frequencies, bandwidths and duration in order to cover a significant portion of the search signal parameter space, as seen in Table \ref{tab:ad-hoc_def}. These simulations are injected at sky locations drawn from a uniform distribution in solid angle over the entire sky. The GA waveforms are all linearly polarised. The SG waveforms are  
circularly polarised, that is, the source is assumed to be optimally oriented. The amplitudes \footnote{The amplitude is represented by $h_{rss}$, defined on page 5} of these simulations are chosen from a grid of maximum strain values given by $(\sqrt{3})^N \times 5 \times 10^{-23}$ with $N$ ranging from 0 to 8.

% elliptically polarised (distributed uniformly in the cosine of the inclination angle of the source) and some are
% The inclination angle of the source is defined as the angle between the total angular momentum vector and the line of sight.
% Injected over hrss with some distribution?

\begin{table}
    \centering
    \begin{tabular}{c c c}
         \hline 
         \hline
          & Gaussian Pulse (GA) & \\
         \hline

          &  & $\tau$ (s) \\
         \hline
 
           & & 0.1   \\
           & & 1   \\
           & & 2.5   \\
           & & 4  \\
         \hline 
          & Sine-Gaussian (SG) & \\
         \hline

          $f_0$ (Hz) & $Q$ &  \\
          \hline
 
          70 &  3 &  \\
          70 &  9 &  \\ 
          70 & 100 &  \\
          100 & 9 &  \\
          153 & 9 &  \\
          235 & 3 &  \\
          235 & 9 &  \\ 
          235 & 100 &  \\
          361 & 9 &  \\
          554 & 9 &  \\
          849 & 3 &  \\
          849 & 9 &  \\
          849 & 100 &  \\ 

         \hline 
          & White Noise Burst (WNB) &  \\ 
         \hline

          $f_{low}$ (Hz) & $\Delta f$ (Hz) & $\tau$ (s) \\ 
          \hline
 
          150 & 100 & 0.1 \\
          300 & 100 & 0.1 \\
          750 & 100 & 0.1 \\
          \hline
          
    \hline
    \hline
         
    \end{tabular}
    \caption{Table of generic ad-hoc simulations with defining parameters used in the O3 all-sky short search.}
    \label{tab:ad-hoc_def}
\end{table}

We also utilise a collection of CCSN waveforms, a population commonly benchmarked in the all-sky short search. The specific mechanisms occurring during the explosion of stars can be very complex, and hence difficult to model. Here, the waveform models cover a variety of mechanisms such as different progenitor star masses, rotation vs non-rotation of progenitors,  explosion type and particular GW signatures. Specifically, we look at 10 neutrino explosion models: Anderson et al. 2017 \cite{Andresen:2017s11} (And s11), M$\ddot{u}$ller et al. 2012 \cite{Muller:2012L15} (Mul L15), Kuroda et al. 2016 \cite{Kuroda:2016SFHx} (Kur SFHx), O’Connor $\&$ Couch 2018 \cite{OConnor:2018-m20} (Oco mesa20), Powell $\&$ M$\ddot{u}$ller 2019 \cite{Powell:2018-s18} (Pow he3.5, s18), Radice et al. 2019 \cite{Radice:2018-s9} (Rad s9, s13, s25), and 1 magnetorotationally-driven explosion model: Abdikamalov et al. 2014 \cite{Abdikamalov:2013sta}.

\subsubsection{Statistic to measure pipeline sensitivity}\label{subsec:hrss}
In order to compare the sensitivities of both GMM post-production approaches, we introduce the root sum square of the GW strain - ie. the $h_{rss}$:

\begin{equation}
    h_{rss} = \sqrt{\int_{-\infty}^{\infty} \Big(h_{+}^2(t) + h_{\times}^2(t) \Big)dt}
    \label{eq:hrss}
\end{equation}

where $h_{+}$ and $h_{\times}$ are the polarisation components of the GW signal. A common way to estimate sensitivities in GW burst searches is to calculate the detection efficiency of given waveforms as a function of $h_{rss}$, which is found by taking the fraction of detected events at a given false alarm threshold over the number of injected events for injected $h_{rss}$ amplitude values.
From $h_{rss}$ in Eqn. \ref{eq:hrss}, we can introduce the $h_{rss50}$ statistic, which corresponds to the $h_{rss}$ amplitude where 50$\%$ detection efficiency is achieved. Since this is a measure of GW strain, smaller $h_{rss50}$ signifies the ability to better detect smaller amplitude signals and hence the improvement of sensitivity. In the results below, we quote the $h_{rss50}$ values per waveform as a way to compare the sensitivities of different methodologies.

\subsection{Comparisons with old GMM method}

In order to directly measure the benefits of our updated methodology, we compare sensitivities for the updated GMM process, here-on referred to as GMM+, to the methodology previously detailed in the paper by D. Lopez {\it et al.} \cite{Lopez_2022}, for the O3a Burst all-sky short search. 

The major changes between these methodologies are:

\begin{itemize}
    \item New signal training set: As mentioned above, the signal GMM is now trained on a set of generic simulated WNBs, well-sampled across the entire short duration, low-frequency parameter space. In previous methodology, the signal GMM was trained on a portion of the ad-hoc waveforms detailed in table \ref{tab:ad-hoc_def}, which sparsely sampled the full signal space and created bias towards this specific set of waveforms.
    \item New method to optimise number of Gaussian components in the models: As demonstrated in Figure \ref{fig:dp_valid_colorbar}, this is now done through the optimisation of detection efficiency at an inverse False Alarm Rate (iFAR) threshold of 50 years over combinations of models with differing number of Gaussian components with validation data. Previously, the Bayesian Information Criterion (BIC) was utilized, however was not reliably favouring models with optimal classification performance.
    \item Removal of the $L_{veto2}$ attribute from the analysis: As it did not hold distinguishable distributions between the signal and noise space, we omitted this attribute.
\end{itemize}

As detailed in \cite{Lopez_2022}, the previous methodology applied a 10\% validation, 70\% training, 20\% testing split to background triggers, and a similar 10\% validation, 70\% training, 20\% testing split to the ad-hoc simulation injections from Table \ref{tab:ad-hoc_def}. For the GMM+ methodology, we retain the same split for background triggers, however apply a 20\% validation, 80\% training split to the generic WNB simulation triggers. Despite this new methodology reserving the entirety of the ad-hoc injection data for testing sensitivities, in the interest of fair comparison we use the same set of 20$\%$ ad-hoc injections to test the sensitivities of both methods in this section.

\begin{figure}
     \centering
     \begin{subfigure}{0.5\textwidth}
         \centering
         \includegraphics[width=\linewidth]{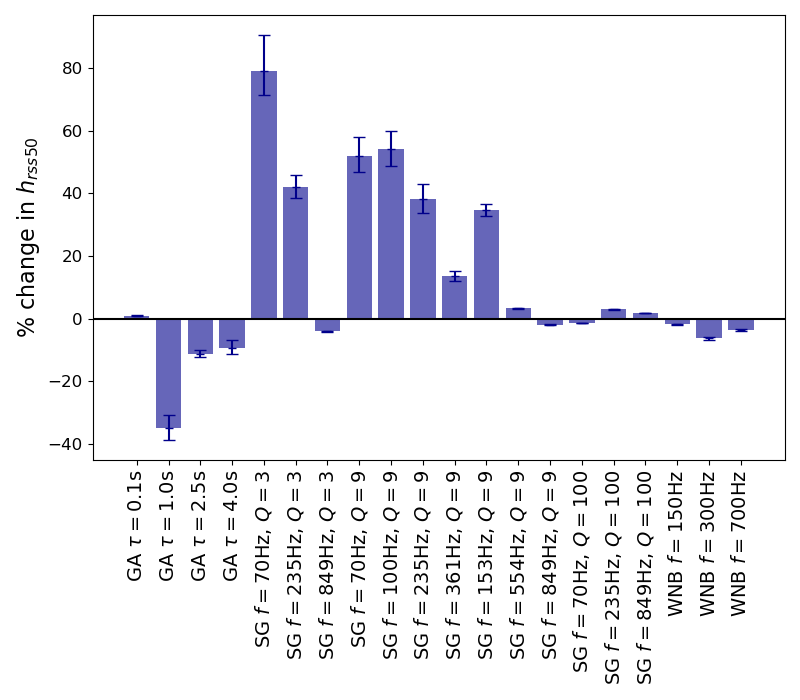}
         \caption{ad-hoc waveforms}
         \label{fig:hrss50_adhoc_O3a}
     \end{subfigure}
     \hfill
     \begin{subfigure}[b]{0.5\textwidth}
         \centering
         \includegraphics[width=\linewidth]{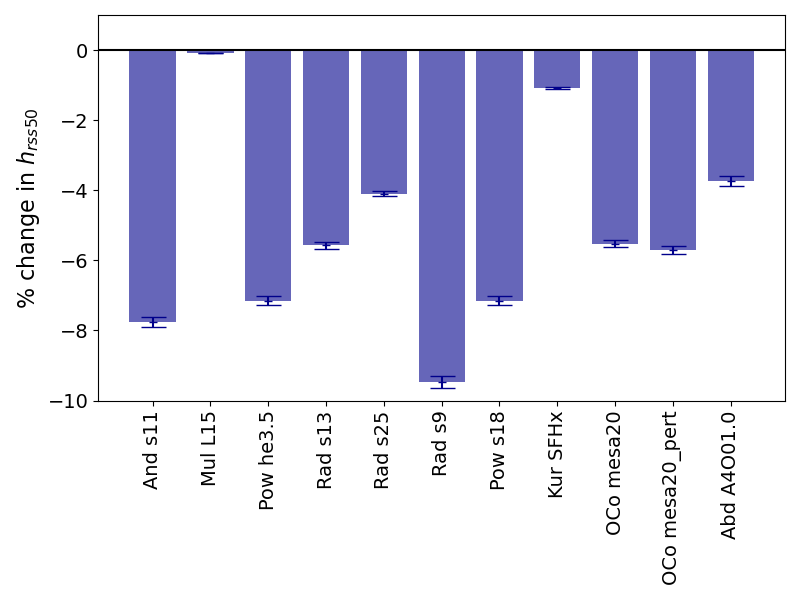}
         \caption{CCSN waveforms}
         \label{fig:hrss50_ccsn_O3a}
     \end{subfigure}
        \caption{Percentage change in $h_{rss50}$ for the updated methodology, GMM+, compared to the old GMM methodology seen in \cite{Lopez_2022} for the generic set of ad-hoc waveform injections and CCSN injections at an iFAR threshold of 100 years. A negative change signifies lower $h_{rss50}$ and hence better sensitivities with the GMM+ methodology.}
        \label{fig:hrss50_comparison_O3a}
\end{figure}

We directly compare the sensitivity of both methodologies to given waveforms by calculating the percentage change in $h_{rss50}$ at iFAR $\geq$ 100 years for GMM+ relative to the previous methodology, as seen in Figure \ref{fig:hrss50_comparison_O3a}. A negative percentage change indicates a lower $h_{rss50}$, and hence an improvement in sensitivity due to the ability to probe smaller amplitude signals for that given waveform. The results for ad-hoc simulation waveforms are shown in Fig \ref{fig:hrss50_adhoc_O3a}. Sensitivities to the majority of Sine-Gaussian waveforms with $Q$=3, $Q$=9 are worsened, however this loss is somewhat expected due to the removal of bias from the models towards these specific injections. Sensitivities to Sine-Gaussian $Q=100$ and White Noise Burst waveforms remain comparable, while GMM+ gains sensitivity to the majority of Gaussian Pulse waveforms. The benefit of eliminating ad-hoc waveform bias from the model is further seen in Figure \ref{fig:hrss50_ccsn_O3a}, where the consistent decrease in $h_{rss50}$ for CCSN injections demonstrates that our new GMM+ methodology improves the pipeline's sensitivity to astrophysical source populations.

\section{Results with the third observing run}
\label{Sec:results_o3b}

We apply the updated methodology of GMM+ to the full third LVK observing run (O3) all-sky short search for the first time. We detail the pipeline's sensitivity to injected burst sources and present GW search results with both the 2-detector LIGOLivingston-LIGOHanford (LH) and 3-detector LIGOLivingston-LIGOHanford-Virgo (LHV) networks. In both cases, models are trained for O3a and second of half of third observation run (O3b) separately to account for difference in detector noise, with sensitivities and search results being combined in the final stage of analysis.

We use the $h_{rss50}$ statistic outlined in section \ref{subsec:hrss} to explore the sensitivity of the cWB+GMM post-production for three populations of waveforms: generic ad-hoc waveforms, astrophysically motivated waveforms from core-collapse supernovae (CCSN) and cosmic string (CS) populations. We use the same set of ad-hoc waveforms detailed in the previous section, consisting of Gaussian Pulse (GA), Sine-Gaussian (SG) and White Noise Burst (WNB) waveforms over a range of frequency, duration and bandwidth properties as detailed in Table \ref{tab:ad-hoc_def}. The same set of CCSN waveforms detailed in Section \ref{Sec:robustness_o3a} are also considered, consisting of 10 neutrino explosion and 1 magnetorotationally-driven explosion models. The final astrophysical population of injections considered is cosmic strings (CS), which are tested by the GMM methodology for the first time. These are one-dimensional topological defects which may form following spontaneous phase transitions in the early universe, with bursts of GW signal expected to be created from both kinks and cusps occurring in the CS loops. From this population we consider 4 waveforms representing CS cusps \cite{Damour:2004CS} with low-frequency cut-off of $1$Hz and high-frequency cut-off of $50$Hz, $150$Hz, $500$Hz, and $1500$Hz, as seen in the O1 LIGO CS search \cite{CosmicString-o1-LVK}. Note that only 10\% of injected CS amplitudes fall inside the analysed frequency band of the algorithm.

Both sensitivities and GW search results are compared to studies completed with other cWB post-production methodologies, namely the cWB standard (STD) post-production detailed in \cite{allskyo3}, and the ML-enhanced decision tree post-production of XGBoost, detailed in \cite{xgboost_2023}.

\subsection{Sensitivities with a 2-detector network}

As mentioned in the above section, for the 2-detector LH network we collect a total coincidence time of $104.9$ days during O3a, with $980.7$ years of background generated. During O3b, there was $101.63$ days of coincident data to be analysed, while time-shifting allowed for $1096.0$ years of background data to be accumulated. 70$\%$ of respective background data was used to train the models, while 10$\%$ is used for validation and 20$\%$ for testing, leaving 196.14 years of background reserved for False Alarm Rate (FAR) calculation in O3a, and 219.2 years in O3b.

\begin{figure*}[t]
    \centering
    \includegraphics[width=\textwidth]{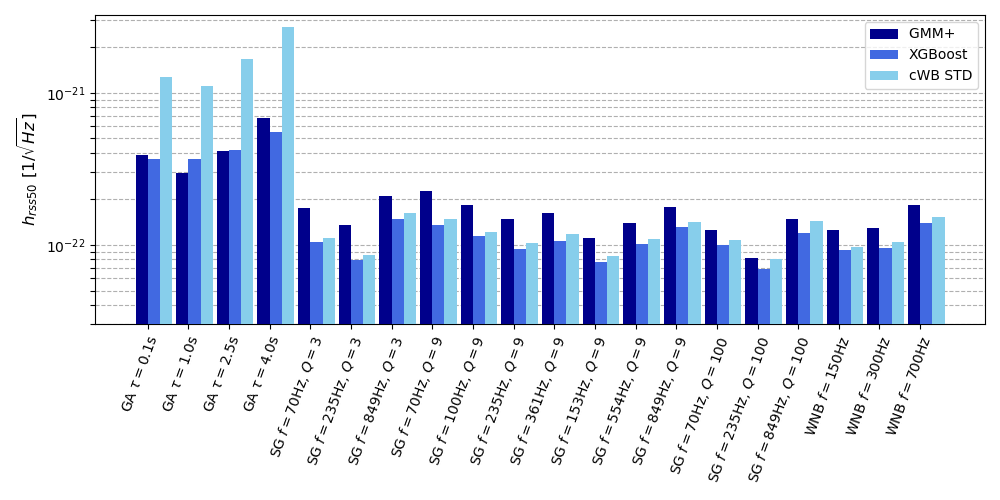}
    \caption{ Sensitivity to generic ad-hoc waveforms detailed in Table \ref{tab:ad-hoc_def} in terms $h_{rss50}$ at iFAR $\geq$ 100 years for the full O3 all-sky short search with 2-detector LH network. Results are shown for GMM+ in dark blue, with comparisons to the XGBoost post-production \cite{xgboost_2023} in royal blue and standard post-production \cite{allskyo3} in light blue. }
    \label{fig:hrss50_adhoc_O3_LH}
\end{figure*}

\begin{figure*}
     \centering
     \begin{subfigure}{0.49\textwidth}
         \centering
         \includegraphics[width=\linewidth]{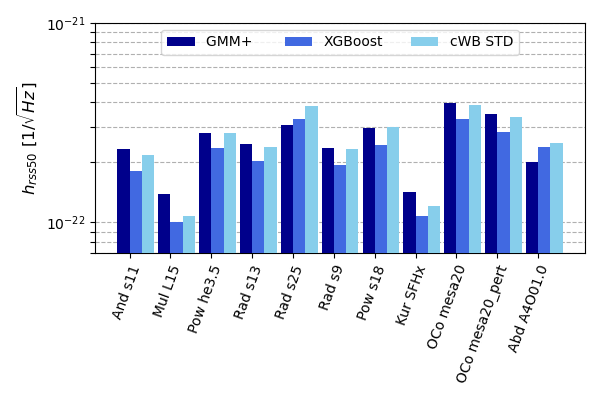}
         \caption{Core-collapse supernovae }
         \label{fig:hrss50_ccsn_O3_LH}
     \end{subfigure}
     \hfill
     \begin{subfigure}[b]{0.49\textwidth}
         \centering
         \includegraphics[width=\linewidth]{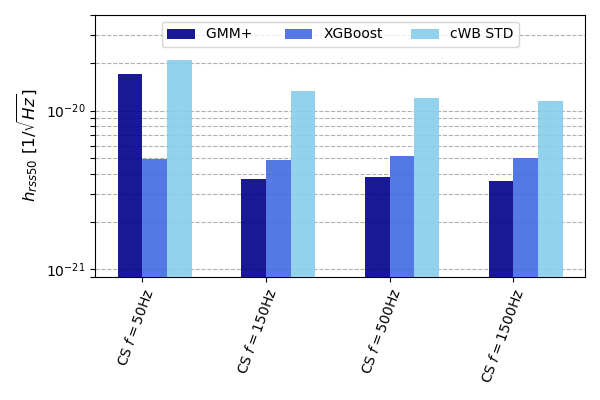}
         \caption{Cosmic strings}
         \label{fig:hrss50_cs_O3_LH}
     \end{subfigure}
        \caption{Sensitivity to astrophysically-motivated waveforms in terms $h_{rss50}$ at iFAR $\geq$ 100 years for the full O3 all-sky short search with 2-detector LH network. Results are shown for GMM+ in dark blue, with comparisons to the XGBoost post-production \cite{xgboost_2023} in royal blue and standard post-production \cite{allskyo3} in light blue.}
        \label{fig:hrss50_astrophysical_O3_LH}
\end{figure*}

% \subsubsection{ad-hoc}

The $h_{rss50}$ sensitivities at iFAR $\geq$ 100 years are quoted for all injected ad-hoc, CCSN and CS injections in Table \ref{tab:hrss50_all}. Figure \ref{fig:hrss50_adhoc_O3_LH} reports the $h_{rss50}$ comparisons for all generic ad-hoc injections detailed in Table \ref{tab:ad-hoc_def} for the 3 cWB post-production methods at a threshold of iFAR $\geq$ 100 years. We see that for Gaussian Pulse waveforms, the GMM+ post-production enhances the sensitivity compared to the Standard post-production (cWB STD) method, while achieving comparable sensitivities to those produced by XGBoost post-production. No sensitivity improvement is seen for sine-Gaussian and white noise burst waveforms, with GMM+ having lower sensitivity than other post-productions by upto $40\%$. The improvement seen for Gaussian pulses demonstrates the ability GMM methodology has to mitigate the effect of blip glitches in the data, as these have previously been one of the most problematic noise source in the low-Q factor parameter space.

% \subsubsection{astrophysical}
%%%%%%%%%%%%%%%%%%%%%%%%%%%%%% LHV SENSITIVITY PLOT - PLACED HERE FOR BETTER PLACEMENT IN PAPER
\begin{figure*}[ht]
    \centering
    \includegraphics[width=\textwidth]{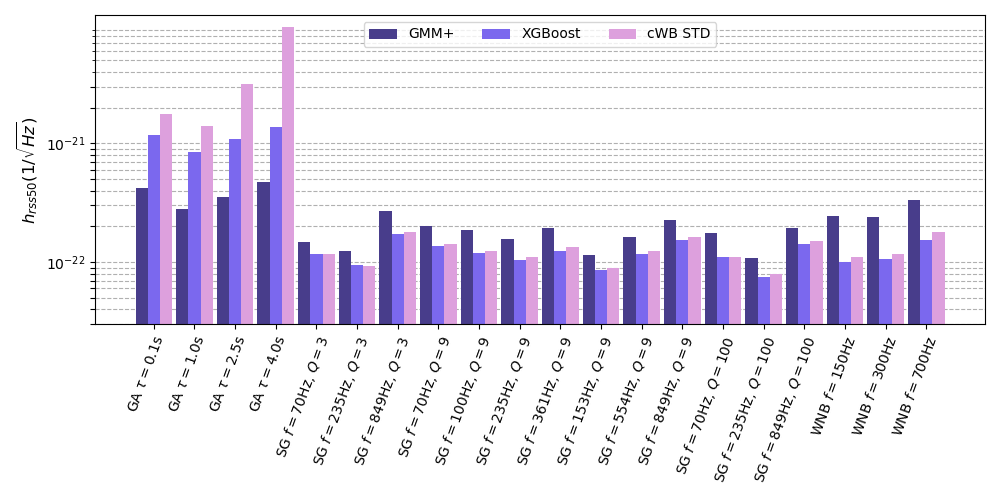}
    \caption{Sensitivity to generic ad-hoc waveforms detailed in Table \ref{tab:ad-hoc_def} in terms $h_{rss50}$ at iFAR $\geq$ 100 years for the full O3 all-sky short search with 3-detector LHV network. Results are shown for GMM+ in dark purple, with comparisons to the XGBoost post-production \cite{xgboost_2023} in medium purple and standard post-production \cite{allskyo3} in light purple.}
    \label{fig:hrss50_adhoc_LHV_o3}
\end{figure*}

\begin{figure*}
     \centering
     \begin{subfigure}{0.49\textwidth}
         \centering
         \includegraphics[width=\linewidth]{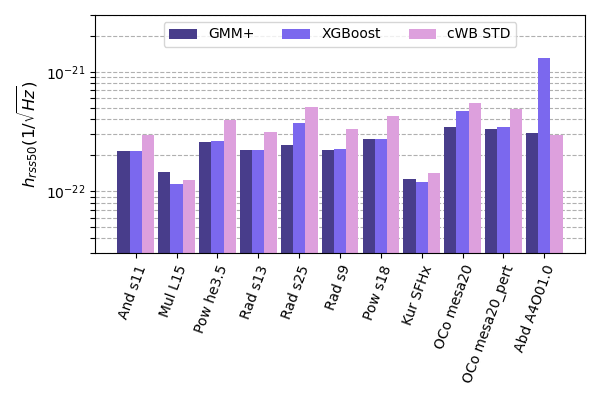}
         \caption{CCSN waveforms}
         \label{fig:hrss50_ccsn_LHV_o3}
     \end{subfigure}
     \hfill
     \begin{subfigure}[b]{0.49\textwidth}
         \centering
         \includegraphics[width=\linewidth]{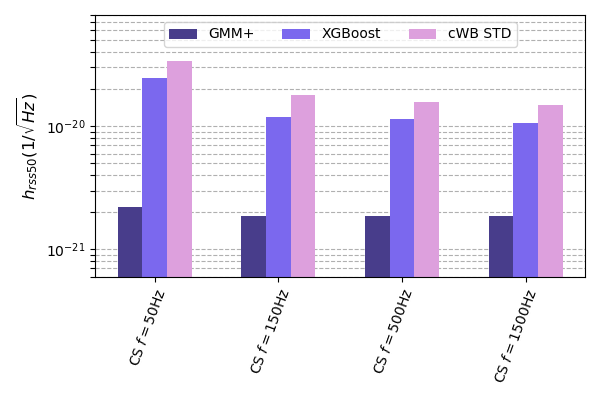}
         \caption{CS waveforms}
         \label{fig:hrss50_cs_LHV_o3}
     \end{subfigure}
        \caption{Sensitivity to astrophysically-motivated waveforms in terms $h_{rss50}$ at iFAR $\geq$ 100 years for the full O3 all-sky short search with 3-detector LHV network. Results are shown for GMM+ in dark purple, with comparisons to the XGBoost post-production \cite{xgboost_2023} in medium purple and standard post-production \cite{allskyo3} in light purple.}
        \label{fig:hrss50_astrophysical_LHV_o3}
\end{figure*}

% Sensitivities to astrophysically-motivated \Sayantan{non-CBC} population injections are reported in , the area in which our is more 

The robustness of the updated methodology to various astrophysically-motivated population injections, especially cosmic strings, is supported by Figure \ref{fig:hrss50_astrophysical_O3_LH}. Here, we again present $h_{rss50}$ estimates at a threshold of iFAR $\geq$ 100 years, with comparisons to the standard \cite{allskyo3} and XGBoost \cite{xgboost_2023} post-productions. Figure \ref{fig:hrss50_ccsn_O3_LH} details the set of CCSN injections. GMM+ observes sensitivity improvements for the {\it Rad s25} \cite{Radice:2018-s9} neutrino-driven explosion model and the magnetorotationally-driven explosion model {\it Abd A4O01.0} \cite{Abdikamalov:2013sta}. The GMM+ methodology still performs well for other waveform models, obtaining sensitivities within 17\% of the standard post-production, and 30\% of XGBoost. 

Sensitivity to cosmic string injections is shown in Figure \ref{fig:hrss50_cs_O3_LH}. The GMM+ post-production brings improvement to all theorised CS sources considered with high-frequency cut-off above 150Hz, with a 30\% reduction to the $h_{rss50}$ seen by XGBoost, and a 75\% reduction to that seen by standard post-production. GMM+ has worse performance for only the CS $f=50$Hz waveform, for which XGBoost achieves the highest sensitivity. The results seen on cosmic string waveforms reinforce that Gaussian Mixture Modelling can effectively mitigate blip glitches and performs well in the low-Q factor parameter space.

\subsection{Sensitivity with a 3-detector network}

We extend the GMM+ post-production to analyse data from the 3-detector LIGOLivingston-LIGOHanford-Virgo (LHV) network, presented here for the first time. For the 3-detector network, we collect 79.12 days of coincident data during O3a, and 72.24 days for O3b. Again, using the time-shifting analysis, we produced 572.9 years of background for O3a and 395.8 years for O3b. In order to have more background data reserved for False Alarm Rate calculation, we altered the background data split for LHV network to be 10\% validation, 60\% training, and 30\% testing, leaving 171.87 years of test data in O3a and 118.75 years in O3b.
The overall methodology remains largely consistent, with minimal changes made to the re-parameterisation of the $N_{ED}$ and $Q_{veto1}$ attributes to account for the dependence of attribute definitions on the number of detectors. These re-parameterisations are seen in Table \ref{tab:attributes}.

% \Jiyoon{The GMM post-production demonstrated better sensitivity in O3a compared to O3b. The O3b background contains a significantly higher rate of noise transients, \note{making it more challenging to distinguish between signals and glitches.} As in LH, the search results are presented for full O3.}

% \subsubsection{ad-hoc}

The $h_{rss50}$ sensitivities at iFAR $\geq$ 100 years are again quoted for all injected ad-hoc, CCSN and CS injections in Table \ref{tab:hrss50_all}, for LHV analysis with GMM+, XGBoost and cWB STD. Figure \ref{fig:hrss50_adhoc_LHV_o3} shows these sensitivities for all generic ad-hoc injections. Here, GMM+ brings improvement to the sensitivity of all Gaussian Pulse waveforms compared to both standard and XGBoost post-production, resulting in an average 85\% and 66\% decrease in $h_{rss50}$, respectively. Similarly to the 2-detector results, GMM+ is less sensitive to sine-Gaussian and white noise burst waveforms by 39\% and 104\% compared to standard post-production, and by 45\% and 126\% compared to XGBoost post-production.

% \subsubsection{astrophysical}
Figure \ref{fig:hrss50_ccsn_LHV_o3} shows the sensitivity of GMM+ post-production on LHV network for CCSN waveforms. The overall sensitivity of GMM+ is comparable to the other methods: compared to standard post-production, GMM+ shows increased sensitivity for 9 out of 11 injected waveforms, with the {\it Mul L15} waveform having slightly lower sensitivity by 16\%, while the {\it Abd A4O01.0} waveform achieves almost the same sensitivity. When compared to XGBoost post-production, GMM+ shows a noticeable improvement in sensitivity for 3 out of 11 injected waveforms, the others except {\it Mul L15} being comparable within 7\%.

Figure \ref{fig:hrss50_cs_LHV_o3} shows the sensitivity of GMM+ post-production on LHV network for cosmic string injections. GMM+ brings considerable improvements in sensitivity for all CS models when compared to both XGBoost and standard post-production, with an average 90\% decrease in $h_{rss50}$ compared to standard post-production and an average 85\% decrease in $h_{rss50}$ compared to XGBoost post-production. This significant improvement in cosmic strings, alongside the good sensitivity the GMM LHV analysis achieves to Gaussian Pulses, further reinforces how well the methodology can mitigate the effect of blip glitches. This is an important result for the 3-detector network, as it has been noted that the LHV network has less efficient discrimination of glitches during trigger production. Unlike the LH network, this arises due to the 3-detector network being sensitive to un-correlated GW polarisations, which means that less glitches can be disregarded based on this information. Through the application of GMM+ as a post-production method, we are able to achieve similar sensitivities to the LH network, effectively discriminating previously problematic glitches.

% \begin{figure*}
%     \centering
%     \includegraphics[width=\textwidth]{}
%     \caption{O3 hrss50 comparisons for GMM+ 2-detector network and 3-detector network. GMM+}
% \end{figure*}

\begin{table*}[p]
    \centering
    \setlength{\tabcolsep}{10pt}
    \begin{tabular}{c | c c c | c c c}
         \hline
         \hline

           & \multicolumn{6}{c}{$h_{rss50}$  $(\times10^{-22}$ $1 / \sqrt{Hz})$} \\
           \hline
          \multirow{2}{*}{Waveform} & \multicolumn{3}{c|}{LH network} & \multicolumn{3}{c}{LHV network}  \\ 
           & GMM+ & STD & XGB & GMM+ & STD & XGB \\
          \hline\hline
          
          {\bf Gaussian Pulse} & & & & & & \\
          $\tau=0.1$ms & 3.9 & 12.6 & 3.6 & 4.2 & 17.5 & 11.7 \\
          $\tau=1.0$ms & 2.9 & 11.1 & 3.7 & 2.8 & 13.9 & 8.4 \\
          $\tau=2.5$ms & 4.2 & 16.7 & 4.2 & 3.5 & 31.8 & 10.9 \\
          $\tau=4.0$ms & 6.8 & 27.0 & 5.5 & 4.7 & 94.5 & 13.8 \\
          \hline
          
          {\bf Sine-Gaussian} & & & & & & \\
          $f_0=70$Hz, Q=3 & 1.8 & 1.1 & 1.0 & 1.5 & 1.2 & 1.2 \\ %circular
          $f_0=70$Hz, Q=9 & 2.2 & 1.5 & 1.4 & 2.0 & 1.4 & 1.4 \\ %linear
          $f_0=70$Hz, Q=100 & 1.3 & 1.1 & 1.0 & 1.8 & 1.1 & 1.0 \\ %circular
          $f_0=100$Hz, Q=9 & 1.8 & 1.2 & 1.1 & 1.9 & 1.2 & 1.2 \\ %linear
          $f_0=153$Hz, Q=9 & 1.1 & 0.8 & 0.8 & 1.1 & 0.9 & 0.9 \\ %circular
          $f_0=235$Hz, Q=3 & 1.3 & 0.9 & 0.8 & 1.2 & 0.9 & 0.9 \\ %circular
          $f_0=235$Hz, Q=9 & 1.5 & 1.0 & 0.9 & 1.6 & 1.1 & 1.0 \\ %linear
          $f_0=235$Hz, Q=100 & 0.8 & 0.8 & 0.7 & 1.1 & 0.8 & 0.7 \\ %circular
          $f_0=361$Hz, Q=9 & 1.6 & 1.2 & 1.1 & 1.9 & 1.3 & 1.2 \\ %linear
          $f_0=554$Hz, Q=9 & 1.4 & 1.1 & 1.0 & 1.6 & 1.2 & 1.2 \\ %circular
          $f_0=849$Hz, Q=3 & 2.1 & 1.6 & 1.5 & 2.7 & 1.8 & 1.7 \\ %circular
          $f_0=849$Hz, Q=9 & 1.8 & 1.4 & 1.3 & 2.3 & 1.6 & 1.5 \\ %circular
          $f_0=849$Hz, Q=100 & 1.5 & 1.4 & 1.2 & 1.9 & 1.5 & 1.4 \\ %circular
          \hline

          % {\bf Sine-Gaussian Q=3} & & & & & & \\
          % $f_0=70$Hz & - & - & - & - & - & - \\
          % $f_0=253$Hz & - & - & - & - & - & - \\
          % $f_0=849$Hz & - & - & - & - & - & - \\
          % {\bf Sine-Gaussian Q=9} & & & & & & \\
          % $f_0=70$Hz & - & - & - & - & - & - \\
          % $f_0=100$Hz & - & - & - & - & - & - \\
          % $f_0=153$Hz & - & - & - & - & - & - \\
          % $f_0=253$Hz & - & - & - & - & - & - \\
          % $f_0=361$Hz & - & - & - & - & - & - \\
          % $f_0=554$Hz & - & - & - & - & - & - \\
          % $f_0=849$Hz & - & - & - & - & - & - \\
          % {\bf Sine-Gaussian Q=100} & & & & & & \\
          % $f_0=70$Hz & - & - & - & - & - & - \\
          % $f_0=253$Hz & - & - & - & - & - & - \\
          % $f_0=849$Hz & - & - & - & - & - & - \\

          {\bf White Noise Burst} & & & & & & \\
          $f_{low}=150$Hz  & 1.2 & 1.0 & 0.9 & 2.4 & 1.1 & 1.0 \\
          $f_{low}=300$Hz & 1.3 & 1.0 & 1.0 & 2.4 & 1.2 & 1.1 \\
          $f_{low}=700$Hz & 1.8 & 1.5 & 1.4 & 3.3 & 1.8 & 1.5 \\
          \hline

          {\bf Core-collapse Supernova} & & & & & & \\
          And s11 & 2.3 & 2.2 & 1.8 & 2.2 & 2.9 & 2.2 \\
          Mul L15 & 1.4 & 1.1 & 1.0 & 1.4 & 1.2 & 1.1 \\
          Pow he3.5 & 2.8 & 2.8 & 2.4 & 2.6 & 3.9 & 2.6 \\
          Rad s13 & 2.5 & 2.4 & 2.0 & 2.2 & 3.1 & 2.2 \\
          Rad s25 & 3.1 & 3.9 & 3.3 & 2.4 & 5.1 & 3.7 \\
          Rad s9 & 2.4 & 1.3 & 1.9 & 2.2 & 3.3 & 1.3 \\
          Pow s18 & 3.0 & 3.0 & 2.4 & 2.7 & 4.2 & 2.7 \\
          Kur SFHx & 1.4 & 1.2 & 1.1 & 1.3 & 1.4 & 1.2 \\
          Oco mesa20 & 4.0 & 3.9 & 3.3 & 3.5 & 5.5 & 4.7 \\
          Oco mesa20\_pert & 3.5 & 3.4 & 2.8 & 3.3 & 4.9 & 3.5 \\
          Abd A4O01.0 & 2.0 & 2.5 & 2.4 & 3.1 & 3.0 & 13.1 \\
          \hline

          {\bf Cosmic String} & & & & & & \\
          $f=50$Hz & 170.0 & 208.3 & 49.8 & 22.1 & 336.4 & 246.7 \\
          $f=150$Hz & 37.0 & 133.5 & 48.8 & 18.7 & 180.2 & 117.9 \\
          $f=500$Hz & 38.1 & 119.6 & 52.0 & 18.7 & 155.6 & 114.0 \\
          $f=1500$Hz & 36.3 & 114.6 & 50.1 & 18.5 & 148.4 & 106.5 \\

         \hline
         \hline
    \end{tabular}
    \caption{Table detailing the $h_{rss50}$ in units of $\times10^{-22}$ $1/\sqrt{Hz}$ achieved at an iFAR $\geq$ 100 years for each injected waveform in O3 across all three cWB post-production methodologies. Values for the STD and XGB post-productions are taken directly from \cite{xgboost_2023}. }
    \label{tab:hrss50_all}
\end{table*}

\subsection{GW detections}

\begin{figure*}[t]
    \centering
    \begin{subfigure}{0.49\textwidth}
        \centering
        \includegraphics[width=\linewidth]{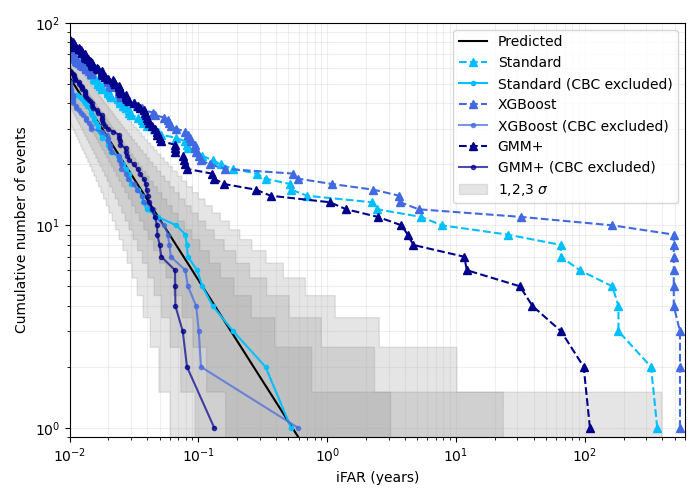}
        \label{fig:openbox-LH}
    \end{subfigure}
    \hfill
    \begin{subfigure}[b]{0.49\textwidth}
        \centering
        \includegraphics[width=\linewidth]{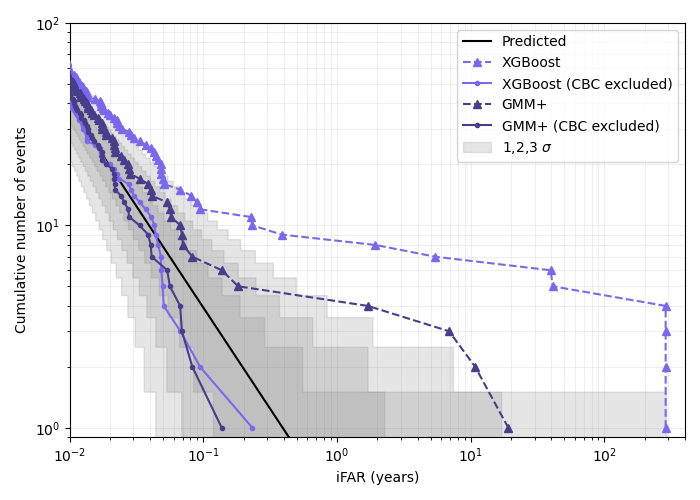}
        \label{fig:openbox-LHV}
    \end{subfigure}
    \caption{Cumulative number of events vs. iFAR found from the O3 all-sky short search for 2-detector LH network ({\it left}) and 3-detector LHV network ({\it right}). Dashed triangular lines represent search results including known CBC detections for the various post-production methods, while solid lines represent the results with known CBC events removed. {\it Left}: search results for LH network, with GMM+ results (dark blue) compared against the XGBoost (royal blue) and standard post-production results (light blue). {\it Right}: search results for LHV network, with GMM+ results (dark purple) compared against XGBoost (medium purple). All post-production methods conclude on null result for non-CBC events.}
    \label{fig:openbox}
\end{figure*}

Here, we discuss the detections in O3 data from the cWB-GMM+ analysis for LH and LHV. All significant events seen are known CBC detections in \cite{GWTC-3}, concluding with a null result on burst-type events. The results are displayed in Figure \ref{fig:openbox}, where the cumulative number of events vs. iFAR is plotted. All search results are notated by the dashed triangular markers, and results with known CBC removed are notated by solid lines. The left-hand plot shows results for the LH network, with GMM+ results compared to the XGBoost and standard cWB post-productions. The GMM+ methodology detects a total of 14 CBC events for iFAR $\geq$ 1 year, similar to the standard analysis (14) and slightly less than XGBoost (16). Overall the GMM+ analysis detects CBCs with lower significance, which is not surprising as the updated methodology has improved sensitivity mainly for the Gaussian pulse and cosmic string waveforms. The loudest event detected by GMM+ with the LH network is GW200224\_222234, with an iFAR close to 110 years. 
%cWB-XGB and cWB-STD have reported this event with iFARs of 548 years and 365 years respectively.
With known CBC events removed, the search results are consistent with the expected background.

The right-hand plot of Figure \ref{fig:openbox} presents the LHV search results, this time only with GMM+ and XGBoost post-productions since the cWB-standard search was not run for LHV network \cite{allskyo3}. For LHV, GMM+ observes a total of 4 events with iFAR $\geq$ 1 year, which is less than detected with the XGBoost post-production (8). The loudest event detected by GMM+ with LHV is GW190412 with an iFAR = 19.10 years. 
%reported by cWB-XGB with iFAR = 40 years. 
Similar to the LH network, the search is consistent with the expected background when known CBCs are removed.

A full breakdown of CBC events detected by cWB-GMM+ is detailed in Appendix  \ref{app:detectedCBC}, with events listed in decreasing order of iFAR and compared to the significance found by standard and XGBoost post-production.

\section{Conclusions}
\label{Sec:conclusions}

Developing searches for un-modeled GW signals from astrophysical systems has always been challenging, due to new classes of noisy transients percolating in any GW signal search that is model-agnostic by nature. cWB has been a back-bone in the un-modelled search for GW transients, however is still affected by noisy transients from the detectors. There are new approaches to post-production with a variety of methods to mitigate the noisy transients. Gaussian Mixture Modelling is one such approach which develops models for signal and noise in the multi-dimensional attribute space under the supervised machine learning framework using the likelihood ratio statistic. While an earlier version of GMM was trained on a suite of ad-hoc waveforms, in this work, we detail the enhancement of the GMM methodology as a post-production to the cWB burst search algorithm. We refer to this enhanced version as GMM+, in which we train our signal models on generic WNB simulations distributed over a broad frequency range, making the search more robust to a wide variety of burst signals. We also improve the validation stage by maximising the detection efficiency over a wide range of numbers of Gaussian components, instead of BIC, which did not reliably select the optimal models. Through updates on the approach to model optimisation, we have removed a bias to ad-hoc waveforms that was previously seen in the GMM all-sky short application, and have demonstrated that the analysis now has increased robustness to a wider class of expected sources within the short GW transient signal parameter space. 

Additionally, from applying the analysis to the gravitational wave data of the third LVK observation run for the first time, we see that both two detector LH and three detector LHV searches can achieve improvement in sensitivities to Gaussian pulses and cosmic strings. The most significant sensitivity improvements are seen within the LHV network, which achieves substantial improvements in sensitivities to Gaussian pulses and cosmic strings compared to other post-production methods, and comparable results to GMM+ analysis with the 2-detector network. It is this improvement in the low Quality factor region which proves the ability GMM+ has in mitigating blip glitches, one of the most problematic classes of noise transients in burst searches. We also obtain comparable sensitivities to those seen by the XGBoost post-production for CCSN. 

The GMM+ post-production detects a similar number of CBC events at iFAR $\geq$ 1 year to the other post-production methods for the LH network, whereas detects half as many for the LHV network. In both cases, GMM+ detects CBC events with less significance than other methods, however this is not the targeted sensitivity space of the search. With known CBC events removed, we conclude with null result for non-CBC events, similarly to the conclusions in \cite{allskyo3,xgboost_2023}. 

{Considering the competitive sensitivities of the GMM and XGBoost post-productions to various astrophysical signals, it may in the future be desirable to run multiple pipelines on a dedicated search for GWs. The combination of multiple pipelines would require the application of a trials factor, however the true implications of this are not yet fully understood and not the target of this work.

The results of GMM+ are encouraging, and there are good prospects to use cWB with GMM+ post-production in burst searches for short gravitational wave transients in future LVK observing runs. The GMM post-production approach is general enough that it can be adapted for any un-modelled search, not specific to cWB. We expect that there will be ongoing efforts of improving the methodology with the future observation runs.

\begin{acknowledgments}
The authors would like to thank Giovanni Prodi for useful comments. 
The authors would like to thank the authors of \cite{xgboost_2023} for the production of cWB triggers used for the analysis in this paper. 
The authors would like to thank Gungwon Kang for interesting discussions.
The authors acknowledge the computational resources which aided the completion of this project, provided by LIGO-Laboratory and support by the National Science Foundation (NSF) Grants No..PHY-0757058 and No.PHY-0823459.
L.S acknowledges support from the Science and Technology Facilities Council [ST/V506692/1 2446745] and the PRIN project 202275HT58. 
SG acknowledges fellowship support from MHRD, Government of India. 
J. S. is supported in part by the National Research Foundation of Korea (NRF) funded by the Ministry of Education (NRF-2022R1I1A207366012).
G.V. acknowledge the support of the National Science Foundation under grant PHY-2207728. 
I.S.H. was supported by Science and Technology Facilities Council (STFC) grants ST/V001736/1 and ST/V005634/1. %(COST)?
AP acknowledges the support from SPARC MoE grant SPARC/2019-2020/P2926/SL, Government of India.
\\
\end{acknowledgments}

\appendix

\section{re-parameterised cWB attributes}\label{app:attributes}
\begin{figure*}
    \centering
    \includegraphics[width=0.99\textwidth]{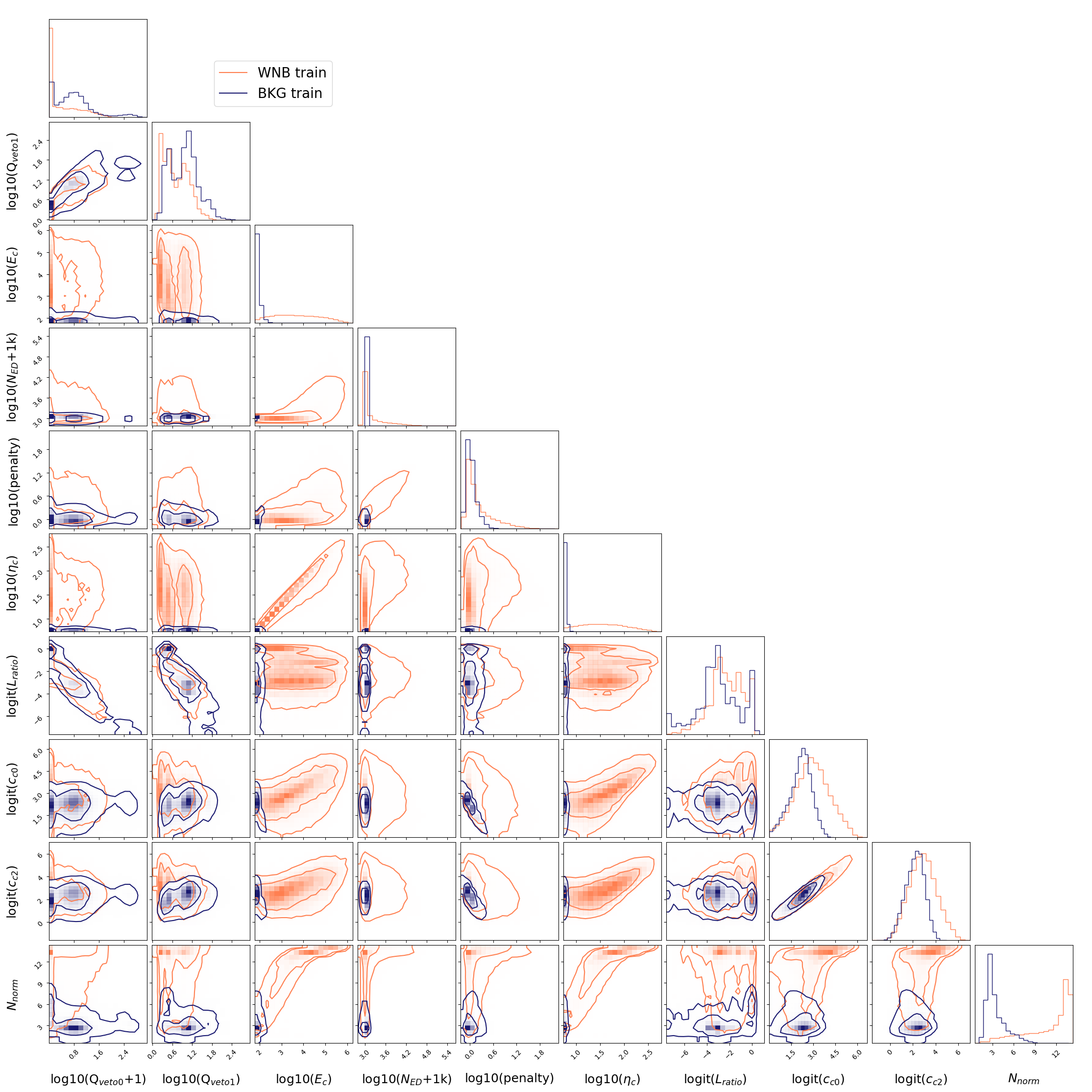}
    \caption{A corner plot showing the distributions of background model training data ({\it BKG train} - navy) and signal model training data ({\it WNB train} - orange) over re-parameterised attributes for the 2-detector LH network. }
    \label{appfig:corner_o3a}
\end{figure*}

As mentioned in section \ref{Sec:Methodology}, the following cWB attributes are considered for the GMM+ analysis: $E_{c}$, $\eta_{c}$, $c_{c0}$, $c_{c2}$, $N_{ED}$, $N_{\mathrm{norm}}$, $penalty$, $Q_{\mathrm{veto0}}$, $Q_{\mathrm{veto1}}$, $L_{\mathrm{veto0}}$, and $L_{\mathrm{veto1}}$. In order to achieve better Gaussian behaviour, these attributes are reparameterised as detailed in Table \ref{tab:attributes}. An example of the distribution of reparameterised attributes is shown for background and signal training data for LH network in Figure \ref{appfig:corner_o3a}, demonstrating that the choice of attributes have distinguishable properties between the 2 populations.
 \\

\section{CBC events detected by cWB-GMM in O3}\label{app:detectedCBC}
Table \ref{combined_table} lists the CBC events detected by GMM+ with iFAR $\geq$ 1 year for both the LH and LHV networks. The iFAR estimates obtained with the other cWB pipelines are listed for the same events. The triggers are obtained from the entire coincident data used for the O3 cWB all-sky short search. 

\begin{table*}
    \centering
    \setlength{\tabcolsep}{8pt}
    \begin{tabular}{c | c c c c | c c c}
         \hline
         \hline

          \multirow{3}{*}{Event Name } & \multicolumn{4}{c|}{LH network} & \multicolumn{3}{c}{LHV network}  \\ 
            & \multirow{2}{*}{T} & GMM+  & STD  & XGB & \multirow{2}{*}{T} & GMM+  & XGB  \\
            &  & iFAR ({\it yr}) & iFAR ({\it yr})  & iFAR ({\it yr}) &  & iFAR ({\it yr}) & iFAR ({\it yr}) \\
          \hline

          GW200224\_222234 & 21.86 & 109.62 & 548.47 & 365.32 & 18.70 & 10.80 & 39.62  \\
          GW190521\_074359 & 32.37 & 98.09 & 490.79 & 326.88 & -5.54 & 0.06 & 286.67 \\ 
          GW190412 & 23.38 & 65.40 & 490.79  & 15.10 & 12.92 & 19.10 & 286.67 \\
          GW190519\_153544 \footnote{For LHV, this event was obtained from the "extended segments" in cWB, for which the minimum analysis segment time is reduced.} & 19.61 & 39.24 & 490.79  & 7.78 & -2.04 & 0.45 & 40.95 \\
          GW191204\_171526 & 13.05 & 31.32 & 32.26  & 91.33 & - & - & - \\
          GW191109\_010717 & 9.63 & 12.18 & 548.47 & 182.66 & - & - & - \\
          GW190828\_063405 & 10.68 & 11.54 & 490.79  & 163.44 & 5.10 & 6.88 & 286.67  \\
          GW190706\_222641 & 7.55 & 4.67 & 490.79   & 65.38 & 0.68 & 1.70 & 286.67  \\
          GW200311\_115853 & 6.35 & 4.22 & 548.47 & 182.66 & -0.24 & 0.03 & 0.39 \\
          GW190521 & 6.49 & 3.77 & 490.79  & 65.38 & -5.54 & 0.06 & 286.67 \\
          GW190408\_181802 & 5.36 & 2.48 & 163.60  & 25.15 & -3.71 & 0.18 & 5.41 \\
          GW191222\_033537 & 3.84 & 1.49 & 5.13 & 2.45 & - & - & - \\
          GW200225\_060421 \footnote{For LH, this event was obtained from the ``extended segments" in cWB, for which the minimum analysis segment time is reduced.} & 3.09 & 1.06 & 5.13 & 2.45 & - & - & - \\
          GW190915\_235702 & 2.96 & 1.05 & 2.28  & 5.36 & -10.55 & 0.004 & 0.019 \\

         \hline
         \hline
    \end{tabular}
      \caption{Table showing the search results of applying cWB-GMM+ to the full O3 data for LH and LHV networks. Only events with iFAR > 1 year are displayed and are arranged in decreasing order of iFAR for the LH network. For the LH network, the iFARs obtained from the cWB-XGBoost (XGB) and cWB Standard (STD) analysis are shown for comparison. For the LHV network, as results are not available for the cWB Standard analysis, cWB-GMM+ results are only compared with cWB-XGB.}
      \label{combined_table}
\end{table*}

\bibliographystyle{apsrev4-1}
\bibliography{references}

\end{document}